# Structure determination of a brownmillerite $Ca_2Co_2O_5$ thin film by Precession Electron Diffraction

P. Boullay, V. Dorcet, O. Pérez, C. Grygiel,

W. Prellier, B. Mercey and M. Hervieu

*Laboratoire CRISMAT, CNRS UMR 6508, ENSICAEN,*

*6 Bd du Maréchal Juin, 14050 Caen, France*

(Dated: 10/02/2009)

**Abstract**

Calcium cobaltite thin films with a nominal ratio Ca/Co=1 were grown on (101)-NdGaO$_3$ substrate by the pulsed laser deposition technique. The structure of the as-deposited metastable phase is solved using a precession electron diffraction 3D dataset recorded from a cross-sectional sample. We found that an ordered oxygen-deficient $Ca_2Co_2O_5$ perovskite of the brownmillerite-type with lattice parameters a ≈ $a_p\sqrt{2}$ ≈ 5.46Å, b ≈ $4a_p$ ≈14.88Å and c ≈ $a_p\sqrt{2}$ ≈ 5.46Å (SG: Ibm2) has been stabilized using the substrate-induced strain. The structure and microstructure of this metastable cobaltite are further discussed and compared to related bulk materials based on our transmission electron microscopy investigations. The present results open the route to the resolution of metastable phases prepared in the form of a thin film using the Precession Electron Diffraction, and should be useful for the solid state chemistry community.

# 1. Introduction

By coupling imaging and analytical capabilities, modern transmission electron microscopes (TEM) afford a unique aptitude for studying materials at a nanometric scale. In terms of structure determination by electron diffraction, several groups have contributed to the field (see [1] as a review) since the pioneering work of B.K. Vainshtein and co-workers [2]. Nonetheless, the strong electron-matter interaction, implying to consider the dynamic theory, has limited the development in materials science laboratories of the structural resolution based on electron diffraction data only. In the Precession Electron Diffraction (PED) method [3], the incident electron beam is rotated along a cone surface, whose axis is the optical axis of the microscope and the vertex is the point of beam incidence on the sample. This diffraction geometry allows integrating the diffracted intensities rods and enlarges the angular resolution while limiting the dynamical scattering effects. This recent diffraction technique is becoming more and more widespread due to the availability of commercial devices that could be installed on most transmission electron microscopes and is already recognized of great interest for accurate symmetry determination of pseudo symmetric structure [4]. The possibility to use PED as a tool for structural investigation has also attracted a growing interest due to its potential application for compounds with large cell parameters that cannot be studied by X-ray powder diffraction (XRPD) or for which XRPD is not sufficient. This is the case for complex zeolite structures that have been solved using PED only [5-7] or in combination with XRPD [8]. Also, the electron diffraction is the only diffraction method available for *ab-initio* structure determination of materials with a small diffracting volume. Since the PED technique has been successfully applied to nanoparticles [9], we decided to use this technique for the structure determination of a metastable phase deposited in the form of a thin film. Compare to nanoparticles, the data collection for a thin film is particularly challenging due to the contribution of the substrate and the specific configuration of the sample.

Among the different thin film techniques, the pulsed laser deposition (PLD) method has been widely used because of its flexibility and versatility, especially for the stabilization of metastable phases and artificial heterostructures that cannot be synthesized using the classical solid state chemistry routes [10]. Several transition metal oxides have thus been prepared including cuprates [11] or manganites [12]. Cobaltite materials represent another class of compounds that have attracted a great interest over recent years due to various potential applications, associated to their structural, chemical and physical properties. These interesting characteristics come from the ability of cobalt to generate mixed valence compounds and to adopt multiple, common or unusual, coordinations especially in oxygen-deficient perovskites. In the pseudo-binary Ca-Co-O system, stable phases have been synthesized for Ca/Co ranging between 1.5 and 0.75: $Ca_3Co_2O_6$ [13] and $Ca_3Co_4O_9$ [14]. Various interesting structural relations have also been observed in $Ca_xCoO_2$ [15], where the $[CoO_2]_n$ layer-spacing varies with the composition. For a ratio Ca/Co=1, an oxygen deficient perovskite $Ca_2Co_2O_5$ has been stabilized [16-18] whose bulk structure is described by alternated stacking of octahedral $[CoO_2]_\infty$ and square-pyramidal $[CoO]_\infty$ layers separated by $[CaO]_\infty$ layers. However, compounds with a ratio Ca/Co=1 can be synthesized only by special processes. Recently, the structure of $Ca_2Co_2O_5$ in the form of textured ceramics prepared by coprecipitation and spark plasma sintering [19] has been reported with a layered structure similar to that of $Ca_3Co_4O_9$ [14]. For this reason, we have investigated the possibilities of stabilizing calcium cobaltite (CCO) with a ratio Ca/Co=1 in the form of thin films using the pulsed laser deposition (PLD) technique. In the present work, the thin films of CCO grown on $NdGaO_3$ (NGO) substrates are investigated by TEM. The structure determination of one of the CCO thin films by PED using a 3D data set recorded on a cross-sectional sample is reported. We found that the as-deposited $Ca_2Co_2O_5$ film is an ordered oxygen-deficient perovskite of the brownmillerite-type whose structure and microstructure are described and compared finally to the related bulk materials.

## 2. Experimental

Targets with a $CaCoO_x$ composition were prepared by solid state reaction. A mixture of $CaCO_3$ and $Co_3O_4$ powders, weighed to the nominal composition was ground in an agate mortar and calcined at 900 °C during 16 h for a full decarbonation. After a slow cooling to room temperature, the powders were crushed again, pressed into a pellet and sintered at 1150°C during 4 h. Thin films were grown by pulsed laser deposition (PLD) on (101)-oriented $NdGaO_3$ substrates (NGO: $a_{NGO}$=5.4979Å, $b_{NGO}$=7.7078Å and $c_{NGO}$=5.4276Å SG n°62 Pnma [20]). Briefly, a pulsed KrF excimer laser (Lambda Physik, Compex, λ=248 nm) was focused onto the target at a fluence of 2.5 J/cm$^2$ with a repetition rate of 2 Hz. The substrate heater was kept at a constant temperature of 600 °C. A background of $O_2$ pressure around 0.05 mbar was applied inside the chamber. After the deposition, the samples were then cooled down to room temperature at a rate of 10 °C/min under 0.05 mbar $O_2$ pressure.

The preliminary structural study was performed by X-ray powder diffraction using a Seifert XRD 3000 diffractometer with Cu $K_{\alpha 1}$ radiation for the θ-2θ measurements. The orientation of the films was done by calibrating the position of the *202* reflection of NGO. Transmission electron microscopy (TEM) was then used for the structural and microstructural characterizations. In a first stage, samples were prepared by scratching the film from the substrate using a diamond tip and the parts of the films were dispersed in alcohol and deposited on a copper grid. The goal was to collect a large part of the film to check its homogeneity over the whole surface and to make easier the complete investigation of the reciprocal space. Cross-section TEM specimens were prepared by the standard techniques using mechanical polishing followed by ion-milling (JEOL Ion-Slicer). The calcium and cobalt contents were measured using Energy-dispersive X-ray spectroscopy (EDS). High Resolution Electron Microscopy (HREM) investigations were performed on a FEI Tecnai G2 30 ($LaB_6$ cathode) microscope operating at 300kV and image simulations were made using the JEMS software. Selected Area Electron Diffraction (SAED) and Precession Electron Diffraction (PED) were recorded on a JEOL 2010 ($LaB_6$ cathode) microscope equipped with a precession module (Spinning Star – Nanomegas). PED data were acquired from several Zone Axes Patterns (ZAP) using the Pleïades (Nanomegas) point detector electrometer that possesses a high dynamic range (24-bits). Intensity extractions were performed using the ELD software and data reduction with Jana2006 [21]. The structure solution was obtained by using the direct methods routine implemented in the SIR2008 software [22].

## 3. Results

*3.1 Epitaxial relations and microstructural features*

A XRPD pattern of a CCO film grown on NGO is given in Fig.1. Besides the reflections from the substrate, a series of six intense diffraction peaks are observed. Considering an epitaxy of the film onto the (101)-NGO substrate, the six reflections can be indexed as the *0k0* reflections with a lattice parameter close to $n \times 7.44$Å, with $n$= 1 or 2 as a function of the conditions limiting the possible reflections (see below). To further analyze the deposited films, observations by transmission electron microscopy were performed on scratched films and a cross-section. The EDS analyses carried out on different films showed that the ratio Ca/Co is equal to 1, within the experimental errors.

The Selected Area Electron Diffraction (SAED) obtained from the cross-sectional sample corresponding to a $[111]_{NGO}$ zone axis pattern (ZAP) is presented in Fig.2a. It results from the superposition of two systems. The first system is associated to the NGO substrate (drawn in full grey lines) with the $[10\text{-}1]^*_{NGO}$ and $[1\text{-}21]^*_{NGO}$ directions corresponding to the out-of-plane and in-plane directions, respectively, with respect to the film/substrate (F/S) interface shown in Fig.2b. A second system (dotted white lines) can be evidenced due to the presence of extra spots. It is associated to the CCO film that exhibits the same in-plane periodicity as NGO along the $[1\text{-}21]^*_{NGO}$ direction ($2 \times d_{1\text{-}21}= 5.456$ Å) but a doubling out-of-plane periodicity as compared to the $[10\text{-}1]^*_{NGO}$ direction ($2 \times d_{10\text{-}1}= 7.726$ Å). In the search of the parametric relations between the NGO substrate and the CCO film, the Fourier Transforms (FT in Fig.2c and 2d) carried out on the two areas indicated in Fig.2b are of importance and evidence the existence of 90° oriented domains in the film. The Fig.2c and 2d, which are the $[001]_{CCO}$ and $[100]_{CCO}$ ZAP, respectively, show similar inter-reticular distances. However, one row of reflections is missing in the $[001]_{CCO}$ ZAP (see arrows in Fig.2c) leading to the conditions limiting the reflection: *0kl, k,l = 2n* and *hk0, h+k = 2n*. The Fig.2a further confirms the epitaxial growth of the film onto NGO. The weak radial splitting along the common $[010]^*_{CCO}$ and $[10\text{-}1]^*_{NGO}$ directions (see encircled spots in Fig.2a) evidences a parametric difference between the out-of-plane periodicity of NGO and CCO of about 2.5%, compatible with the results obtained by XRPD. This value estimated from the whole thickness of the film results from the evolution of the cell parameters with the film thickness (see Fig. 3 and below). Indeed, the FT from small areas at the interface indicate almost no difference between the strained film and the substrate, notably for the in-plane parameters. From FT obtained in the thick part of the film, a difference is observed between $c_{CCO}$ and $a_{CCO}$, with $c_{CCO} > a_{CCO}$, associated to the film relaxation. For the structural study of the CCO film one would further consider an orthorhombic lattice with the following cell parameters: $a_{CCO} \approx a_p\sqrt{2} \approx 5.46$Å, $b_{CCO} \approx 4a_p \approx 14.88$Å and $c_{CCO} \approx a_p\sqrt{2} \approx 5.46$Å. Numerical values are taken from the in-plane periodicity of NGO for $a_{CCO} / c_{CCO}$ (constrained to the same value) and from the out-of-plane periodicity measured by XRPD for $b_{CCO}$. These parameters are consistent with those obtained in the Ga and Al doped cobaltites [23, 24].

The cross section multi-beam low magnification image (Fig.3) confirms the good quality of the as-deposited CCO film with a thickness of about 130 nm. No secondary phase has been detected at the interface or in the film, and the interface is well-defined and coherent. At the interface (Figs. 2b and 3), intensive strain field in the form of modulations of the contrast are locally observed, which can result of the misfit accommodation through the formation of local distortions, antiphase boundaries or dislocations generating Moiré patterns (see [25] and references therein). The Fig.3 also evidences a nucleation of CCO grains through competitive growth resulting in a columnar morphology. The width of the vertical columns is approximately 60 nm. The thicker part of the film is marked by a weak strain contrast, in the form of crossed bands of few nanometers wide. In the bottom left part of the Fig.3, the FT obtained from one of these areas shows the shape of the spots, which exhibit elongated arms along directions close to $[130]^*_{CCO}$ and $[1\bar{3}0]^*_{CCO}$. This suggests a metastable state, intermediate between the tweed structure and the micro-twinning (MT).

*3.2 Structure determination using electron diffraction*

The most typical SAED patterns of the CCO cross-section sample, obtained by a tilting experiment around $[010]^*$, are shown Fig.4. The brightest reflections can be indexed in a body-centred orthorhombic unit cell using the parameters determined above with the conditions limiting the reflections *hkl*: $h+k+l=2n$ and *0kl*: $k=2n$ $(l=2n)$ and the compatible space groups Ibmm (74:cab) and Ibm2 (46:ba-c). Weaker spots show up in the [102] ZAP and can be separated in two different sets (Fig.4c). The first set (see encircled spots) is related to the NGO substrate while the second set consisting of extra rows (white arrows in Fig.4c) is related to the existence of a superstructure as referred to the above-mentioned cell. At this point of the structure determination, this information is eluded and will be discussed later on.

Precession Electron Diffraction (PED) patterns were recorded for several ZAP at a precession semi-angle of 1.5° (~26mrad) with a high sensitive 24-bits point detector electrometer (PLEIADES Nanomegas). The area for the data collection was chosen to avoid as much as possible the contribution from the substrate and from different twin domains. In Fig.5, the recorded data look like images obtained with a CCD camera or an image plate, but they actually correspond to the scans of the PED patterns on the point detector electrometer with a scan step of 8 pixels (about 0.012 Å$^{-1}$) and a step time of 0.01s (1 revolution of electron beam around the microscope axis). Due to its high dynamic range combined with the fact that the electrometer aperture size is larger than the diffracted beams, the beam can enter entirely in the electrometer and the integrated intensity of the diffracted beam can be measured directly. The values of the integrated intensities are extracted using ELD and, after indexing, an *h k l intensity* file is obtained for each zone axis. From the recorded patterns, different artefacts are present such as negative intensities or non-zero intensities for kinematically forbidden reflections (indicating the presence of residual dynamical effects). This is particularly observed for the spots close to the transmitted beam which leads us to remove the reflections below g=0.4 Å$^{-1}$ (g=1/d$_{hkl}$) for all the recorded patterns. An upper limit was also applied and the reflections with g>1.6 Å$^{-1}$ were excluded considering both the possibilities of ZOLZ (Zero Order Laue Zone) and HOLZ (High Order Laue Zone) overlapping and partial integration of the diffraction rods at higher g values. The g-restricted *h k l intensity* files were merged with Jana2006 using the common [010]* row of reflections and data reduction was performed considering the Ibm2 and Ibmm space groups that were both tested for structural solution in SIR2008 using the input parameters given in Table 1. The best solution was obtained for the Ibm2 space group with a R value close to 25% and corresponds to the structure presented Fig.6 that reveals three cationic positions and five oxygen positions (see atomic positions in Table 1). One clearly recognizes a structure of the brownmillerite-type [26] with two extra oxygen positions (O2 and O5) that can be eliminated after examination of the interatomic distances for O5 (Ca-O5 in Table 2) and the Co2 coordination polyhedron for O2 (O3 is the only choice to get a proper tetrahedral environment). Based on this structure obtained by SIR, HREM image simulations were performed and compared with the experimental images. We actually take advantage of the existence of twins to locate a zone where both [100] and [001] orientations are present. The result displayed in Fig.7, with a focus (Δf) close to -20nm and crystal thickness (t) to 6nm, shows a fairly good agreement for both orientations.

## 4. Discussion

The question of the improvement of the as-obtained structure using a classical refinement procedure should be addressed. In this aim, electron scattering factor constants were introduced for Ca, Co and O atoms in Jana2006 [21] refinement program as coefficients for analytical approximation according to the ones used in SIR2008 [22]. The model previously determined with SIR2008 (O2 and O5 oxygen positions were not considered) has been used in Jana2006. Isotropic atomic displacement parameters equal to 0.005 Å$^2$ were considered for all the atoms. The refinement of the scale factor leads to an agreement factor based on the structure factor equal to R$_F$=28.15%. To take into account the 90° oriented domains observed in Fig.2b, the twin law (-z,y,x) corresponding to a 90° rotation around the **b** axis has been introduced. Different twin fractions were then tested for the two domains. The best $R_F$ value is obtained for a 80/20 distribution of domains ($R_F$=27.12%). Nevertheless the refinement modifies drastically the different atomic parameters (x,y,z, $u_{iso}$) (mainly for the O atoms) but does not improve the reliability factor. These modifications induce changes in the cationic environments and, in particular, the coordination of Co2 does not fit any more with the expected ideal tetrahedron. Furthermore, the majority of the $u_{iso}$ reaches negative values.

To analyse this result, Fourier observed maps are calculated using the phases and the observed structure factors considering the initial model ($R_F$=27.12%) before refinement. The resulting maps exhibit a strong noise level. The location of the cations is relatively clear even if they are surrounded by significant electron density peaks (see Fig. 8a, b and d). The position of the oxygen atoms is more difficult to define and local maxima in the electron density can be affected to the extra oxygen positions as determined by SIR2008 (Table 1). In the Fig. 8a, the second maximum corresponds to the expected position of the O1 atoms but a strong elongation along [101] of the electron density is also revealed. The same type of elongated electron density is observed in the figure 8d and leads to the existence of a false maximum that SIR2008 identifies as the O5 oxygen position. In this case the expected oxygen atom (O4) is actually located at another y level (see fig. 8c). Similarly at y=0.75, the existence of a local maximum (O2 from SIR2008) leads to a solution where Co2 is located out of its coordination polyhedron. All these features clearly illustrate the problems of convergence observed during the previous refinement procedure. They can be attributed to residual dynamical scattering, radiation damage generating a decay of the sample, the sparseness of the data set and the applied g selection that notably excluded strong, low order reflections. Therefore the maximum-entropy method (MEM) provides an interesting alternative to the conventional inverse Fourier transform since it reduces notably the truncation effects [27-30]. Accurate electron density distribution was achieved from MEM calculation using the program BAYMEM [31]; input data correspond to experimental phased structure factors and the normalization is ensured by adding the total number of electrons in the unit cell *i.e.* F(0 0 0) =568. The unit cell was divided into a grid of 72 x 216 x 72 pixels. The reliability factor of the MEM, $R_{MEM} = \sum|F_{obs} - F_{MEM}|/\sum|F_{obs}|$ =0.053, where $F_{MEM}$ is the structure factor calculated from the electron density obtained by the MEM. The figures 9a, b, c and d show the improvement of the electron density. The positions of the different atoms are clarified. The noise level strongly dropped and the lines of density almost disappeared. O1 and O3 can be unambiguously identified and a splitting of the O4 atom can be suggested following the electron density observed in the Fig. 9c.

The existence of a local maximum (O2) or the splitting of the O4 atom cannot only be considered as artefacts due to the sparseness of the data set. They can be related to structural peculiarities often encountered in brownmillerite-type compounds and generally attributed to two main chemical and structural mechanisms, namely the ability to adapt extra oxygens in the framework, depending on the possible oxidation states of the cations (cobalt is a perfect candidate) and the different orientations of the tetrahedral chains. The latter mechanism is fully consistent with the existence of the additional diffraction spots previously evidenced in Fig. 4c and enlightened by white arrows. By tilting around [010]*, these spots are clearly observed in the $[102]_{CCO}$ SAED patterns (superimposed to $[201]_{CCO}$ due to twin domains). They are similar to those previously reported in the brownmillerite $Sr_2MnGaO_5$ [32] and associated to the existence of a superstructure described as a modulated structure, with a modulation vector **q**=1/2**c*** and the super-space group Imma(00γ)s00. In the present case, considering the Ibm2 average cell, the modulation vector associated to this superstructure would be **q**=1/2**a***. The origin of this superstructure is associated to ordering of the two mirror orientations of the tetrahedral chains, commonly denoted "left" (L) and "right" (R) [33].

A second peculiarity of the film is frequently observed in the [001]-oriented domains and consists in local modulations of the contrast. The as-formed nanodomains (see illustration in Fig.10) are a few unit cells wide and the HREM image shows a twin-free area, without rupture of the perovskite-type layers. The corresponding FT (inset in Fig.10) exhibits additional reflections with non-integer values of $h$. In the majority of the domains[1], the satellites can be indexed using a modulation vector in the form **q**=α**a*** (assigned to the basic non-extinct reflections) where the α values range between 2/3 and 1/2 (in the Fig.10, α is about 0.56). Based on crystallographic and chemical considerations on the brownmillerite compounds, two probable origins can be proposed for these nanodomains, which are likely associated to short range ordering (SRO). In the $Ca_2M_2O_5$, modulation vectors **q**=2/3**c*** and 5/8**c*** (SSG: Imma(00γ)s00) have already been reported in an aluminium doped cobaltite [24] and associated to the ordering of the L and R chains of tetrahedra. Similar phenomena have been reported in the high temperature forms of the brownmillerite $Ca_2Fe_2O_5$, beyond the Pnma to Imma transition [35] with a γ value of 0.588 (SSG: Imma(00γ)s00) and in $Ca_2Al_2O_5$ with γ=0.595 [36] (SSG: Imma(00γ)s00). In the present case (nanodomains), they would be associated to a local ordering different from the ones observed in the [102]$_{CCO}$ SAED patterns (Fig. 4c). A second possible origin could be associated to local variation of the oxygen content. If one admits that the formulation of the as-deposited film is close to Ca$_2$Co$_2$O$_5$, the brownmillerite can take up during the oxidation process a small amount of oxygen, through the [001] tunnels of the [CoO]$_\infty$ layers. They transform the Co tetrahedra in pyramids and/or octahedra, in a more or less ordered way whose mechanism has been observed in numerous ABO$_{2.5+x}$ compounds [34, 37-39], which exhibit long and short range ordering phenomena.

---

[1] Note that in some domains (less frequent and not shown), a component along **b*** is also observed, as in the brownmillerite $Sr_2Fe_2O_5$ [34].

Referring to the previous work on a Ca$_2$Co$_2$O$_5$ bulk compound [16], built up from CoO$_5$ pyramids, it appears that the two materials, CCO$_{film}$ and CCO$_{bulk}$, crystallize in two different structures with different cell parameters and symmetry. Nonetheless considering the perovskite subcell, the two compounds have very close cell parameters and the same volume cell per cobalt (55.8 Å$^3$ [16] vs. 55.5 Å$^3$ [this work]) in agreement with the fact that the oxygen content and expected oxidation state of the cobalt are essentially similar. Besides the specific characteristics of films growth, a small amount of additional oxygen could also favour the stabilisation of a brownmillerite-type structure, as observed in the bulk materials by doping both the A and B sites [40, 41]. Considering the c/a ratio commonly observed for the Ca$_2$M$_2$O$_5$ brownmillerite compounds [24, 35, 36], the accommodation of the parametric difference is ensured through the columnar morphology of the film and microtwinning. The presence of superstructures and nanodomains, associated to aperiodic sequences of different configurations of the tetrahedral chains, suggests that the distortion of these chains could be one of the structural mechanisms for strain relaxation in this Ca$_2$Co$_2$O$_5$ brownmillerite-type film.

## 5. Conclusion

A well-crystallized epitaxial Ca$_2$Co$_2$O$_5$ thin film has been deposited on NdGaO$_3$ substrate by the pulsed laser deposition technique. Its average brownmillerite structure (SG: Ibm2 with a≈a$_p$√2, b≈4a$_p$ and c≈a$_p$√2) has been determined by direct methods (SIR2008) using precession electron diffraction data recorded on a cross-sectional sample. The impossibility to refine the model obtained by direct methods can be attributed to the sparseness of the dataset and the dynamical scattering effects still present in our data extracted from low indexes zone axis PED patterns. The use of PED for structural resolution needs further developments in terms of structure determination software capable to account for the precession geometry and dynamical scattering (as in [42]), at least, considering 2-beams interactions. Nevertheless, such report opens the route to the resolution of metastable phases prepared in the form of a thin film using the Precession Electron Diffraction, and should be useful for the solid state chemistry community.

## Acknowledgments

This work is carried out in the frame of the SONDE project supported by the ANR (ANR-06-BLAN-0331). Dr. V. Dorcet acknowledges the council of the region Basse-Normandie for financial support.

## $Ca_2Co_2O_5$

| | | | |
|---|---|---|---|
| space group | Ibm2 (46 : ba-c) | | |
| ind. obs. ref. | 255 | | |
| cell parameters (Å) | 5.46 | 14.88 | 5.46 |
| cell content | 8 Ca | 8 Co | 20 O |

**structure results**

| Atom | x | y | z | U (Å$^2$) |
|---|---|---|---|---|
| Ca1 | 0.992 | 0.890 | 0.585 | 0.0005 |
| Co1 | 0.000 | 0.000 | 0.129 | 0.0013 |
| Co2 | 0.039 | 0.750 | 0.077 | 0.0003 |
| O1 | 0.277 | 0.993 | 0.412 | 0.0100 |
| *[ O2* | *0.787* | *0.750* | *0.276* | *0.1194 ]* |
| O3 | 0.180 | 0.750 | 0.298 | 0.0855 |
| O4 | 0.932 | 0.860 | 0.065 | 0.0150 |
| *[ O5* | *0.855* | *0.891* | *0.423* | *0.0022 ]* |

**Table 1**: Fractional coordinates and thermal parameters as obtained by direct methods (SIR2008) based on the PED data recorded on the CCO film. The atomic positions are listed and numbered following the SIR2008 output for the best trial (R~25%). Two oxygen positions (O2 and O5 in italic) were removed for the retained structure solution.

| Atom1 | Atom2 | $d_{1-2}$ (Å) | |
|---|---|---|---|
| Co1 | O1 | 1.701 | ×2 |
|  | O4 | 2.142 | ×2 |
|  | O1 | 2.166 | ×2 |
|  | *O5* | *2.415* | *×2* |
|  | *O5* | *2.770* | *×2* |
| Co2 | O3 | 1.437 |  |
|  | O4 | 1.740 | ×2 |
|  | *O2* | *1.753* |  |
|  | O3 | 2.159 |  |
|  | *O2* | *2.420* |  |
|  | *O5* | *2.994* | *×2* |
| Ca1 | *O5* | *1.157* |  |
|  | O4 | 2.360 |  |
|  | O1 | 2.377 |  |
|  | O1 | 2.462 |  |
|  | *O5* | *2.644* |  |
|  | O1 | 2.666 |  |
|  | O4 | 2.681 |  |
|  | O1 | 2.752 |  |
|  | *O2* | *2.785* |  |
|  | O3 | 2.806 |  |
|  | O4 | 2.891 |  |
|  | *O2* | *2.910* |  |
|  | O3 | 2.989 |  |

**Table 2**: Main interatomic distances in the coordination sphere (<3Å) of Ca and Co atomic positions obtained from SIR2008. O2 and O5 oxygen positions not retained for the final model are included.

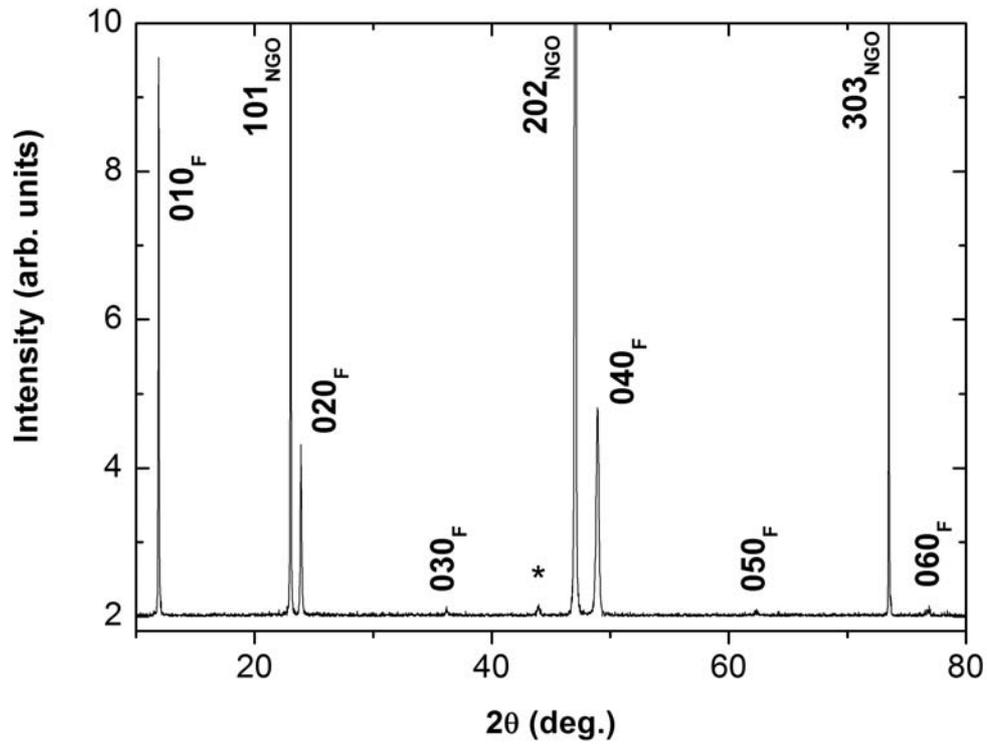

**Figure 1**: Typical XRPD pattern of a calcium cobaltite thin film (Ca/Co=1) grown on NdGaO$_3$ under optimized conditions: p(O$_2$)=0.05 mbar, T=600°C, E$_{laser}$=200 mJ. Indexes F and NGO denote the film and the NdGaO$_3$, respectively. The peak labelled with a star is generated by the sample holder.

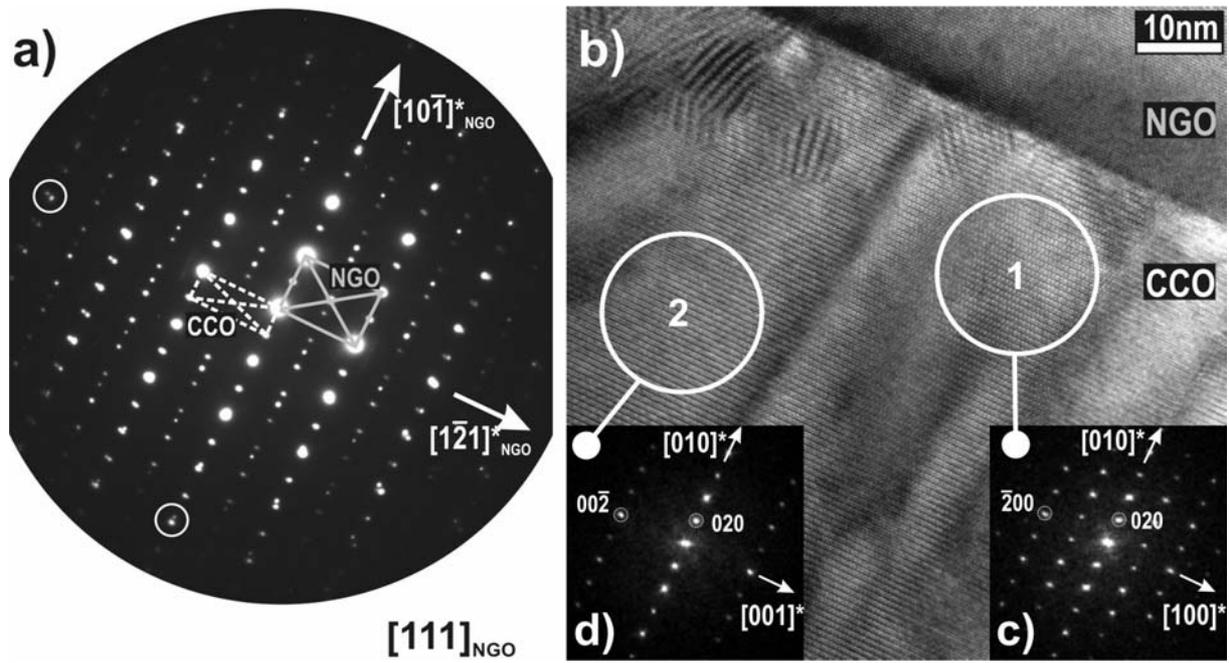

**Figure 2**: In a), $[111]_{NGO}$ ZAP resulting from the superposition of reflections from the NGO substrate (motif in full gray lines) and the epitaxially grown CCO film (motif in dotted white lines). In b), the corresponding image shows the film/substrate interface and reveals the existence of 90° oriented domains in the CCO film (see FT in inset from the areas 1 and 2).

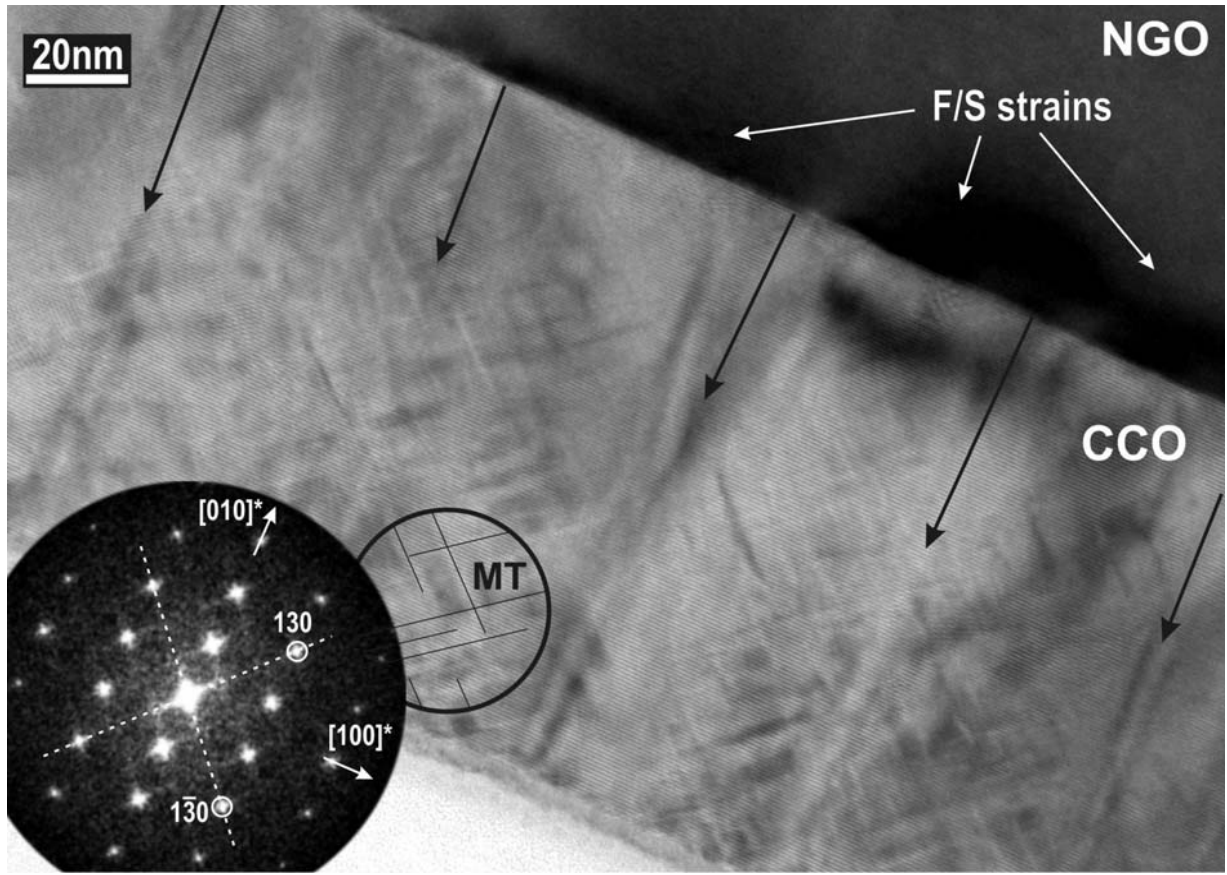

**Figure 3**: Multi-beam low magnification image taken from a cross sectional sample revealing characteristic microstructural features. The CCO film presents a columnar growth morphology (indicated by black arrows as a guide to the eyes). The accommodation of stresses is evidence through contrast modulations and Moiré patterns at the film/substrate (F/S) interface or by micro-twinning (MT) as the film thickness increases.

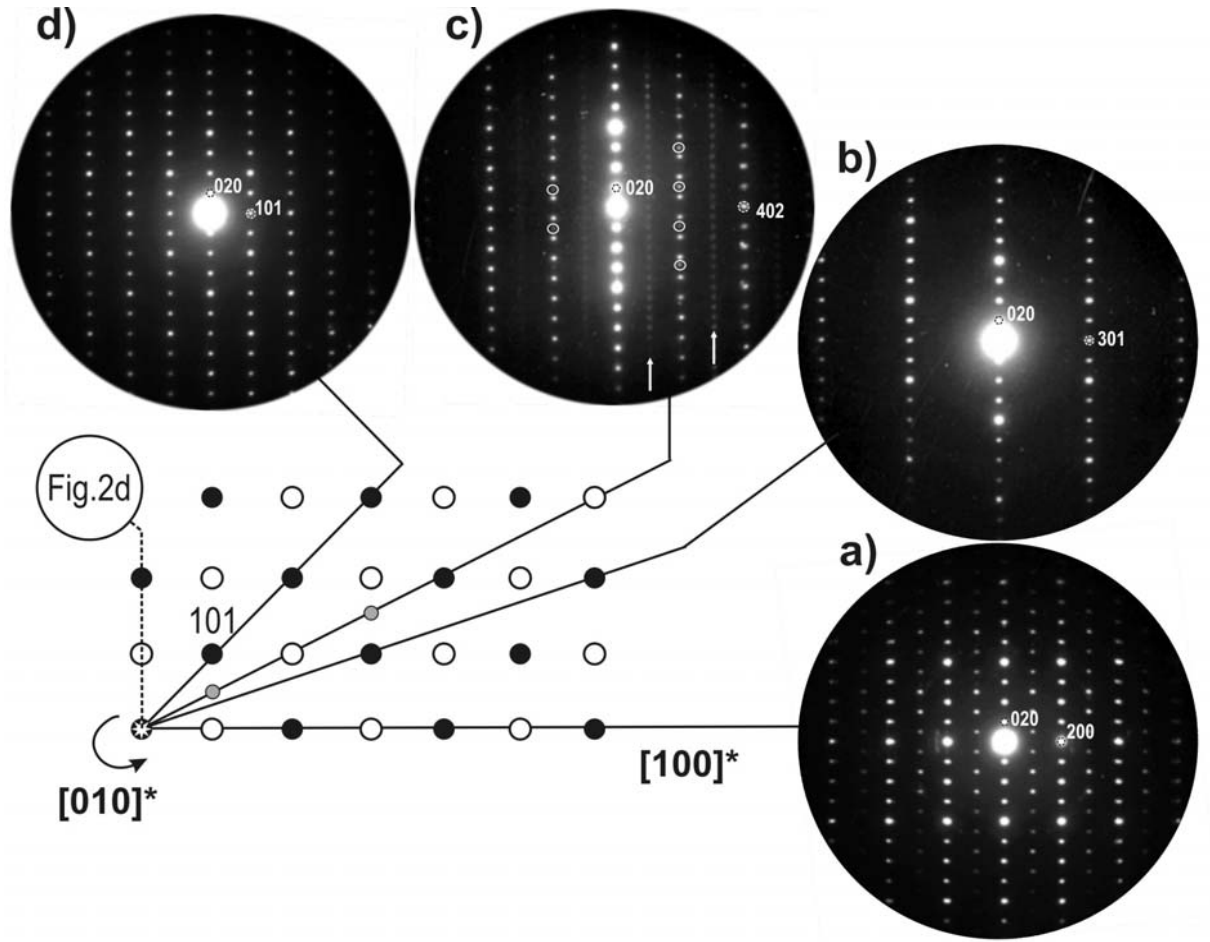

**Figure 4**: SAED patterns obtained by a tilting experiment around [010]* from the CCO film. A schematic drawing of the reconstructed [100] zone axis pattern is given in order to visualize the angular relations between each orientation. In this scheme dark spots correspond to the reciprocal plane (h0l)*, white spots to the (h1l)* plane and the two gray spots indicate the position of the extra rows of reflections observed in c). The main set of reflections is compatible with the space groups Ibmm (74:cab) and Ibm2 (46:ba-c) considering $a \approx a_p \sqrt{2}$, $b \approx 4a_p$ and $c \approx a_p \sqrt{2}$. The four zone axes patterns correspond to a) [001], b) [10-3], c) [10-2] and d) [10-1]. In c) the rows of weak spots indicated by white arrows can be related to the existence of a superstructure similar to the one reported in [32].

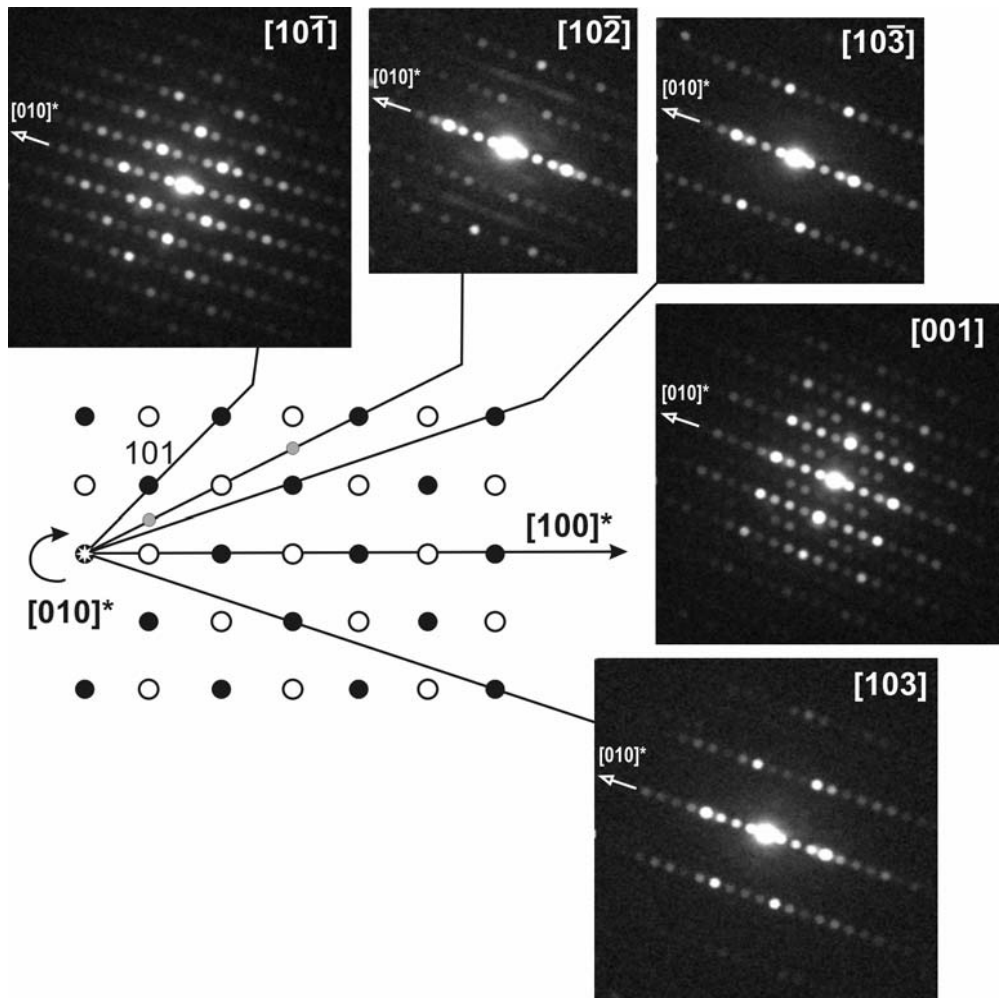

**Figure 5**: These images are obtained by the scanning of the PED patterns on our point detector electrometer. The intensities present on these five sections of the reciprocal space have been extracted for the structure determination. They have the [010]* direction in common that is used for scaling during the merging of the extracted 2D datasets. A schematic drawing of the [100] zone axis pattern is given in order to visualize the angular relations between each orientation. In this scheme dark spots correspond to the reciprocal plane (h0l)*, white spots to the (h1l)* plane and the two gray spots indicate the position of the extra rows of reflections observed in the [10-2] ZAP. For the three larger images the acquisition time is about 11 minutes.

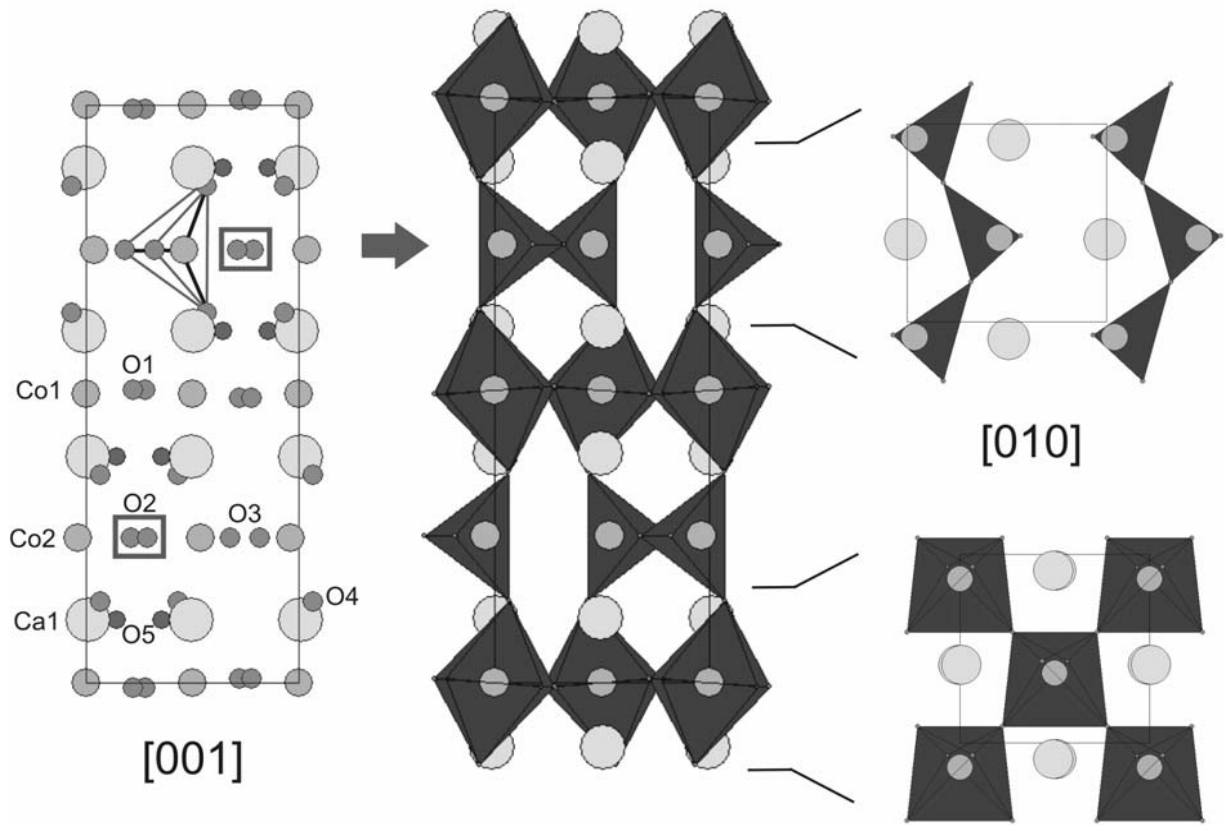

**Figure 6**: Structure solution as obtained using the direct methods of SIR2008 (see Table 1 for atomic positions and Table 2 for inter-atomic distances). After the removal of the two extra oxygen positions (O2 and O5) based on solid state chemistry considerations, a brownmillerite-type structure is obtained.

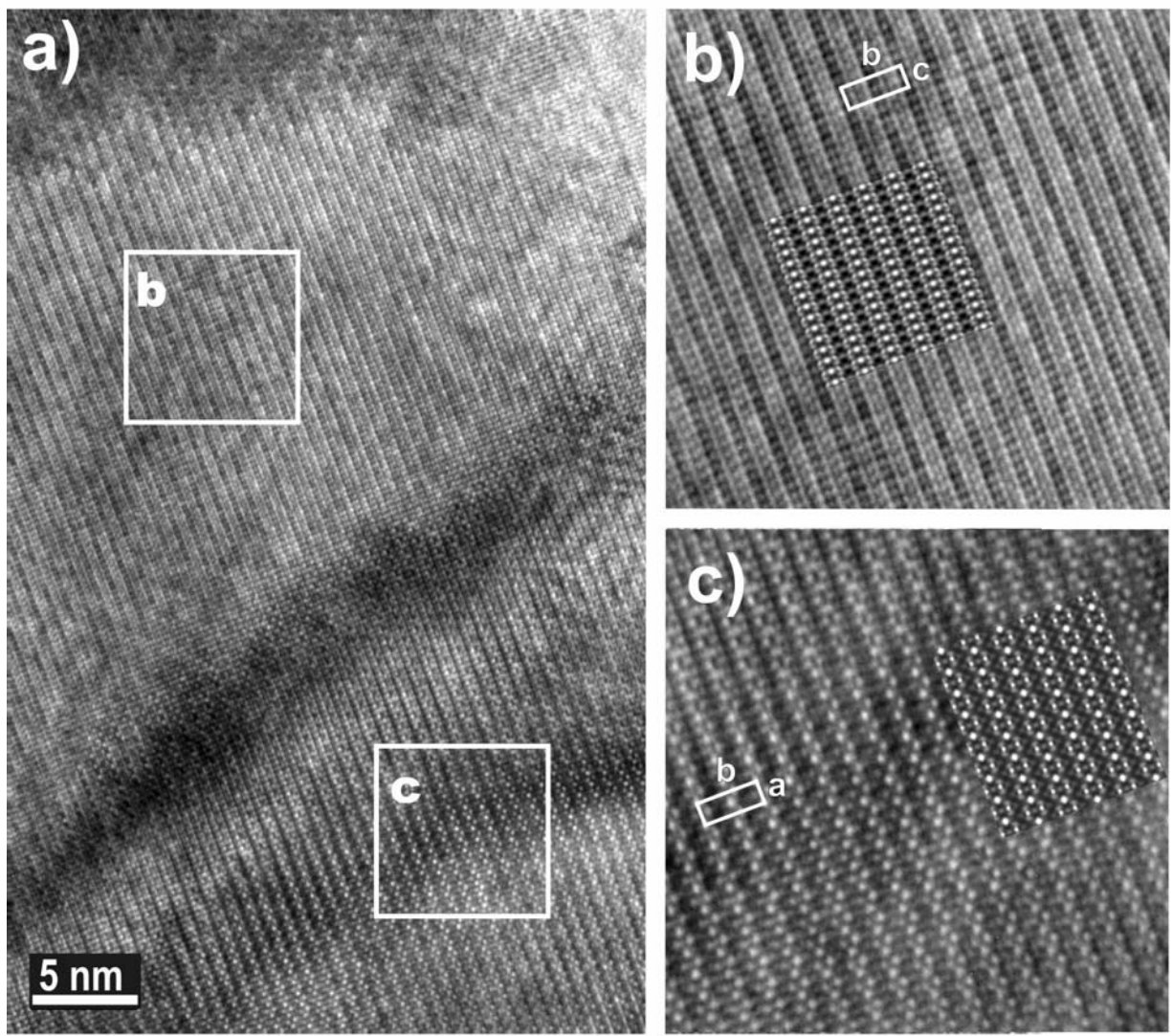

**Figure 7**: In a), HREM image of a part of the film where 90° oriented domains are present. Areas marked by a square are enlarged in b) and c) and correspond, respectively, to [100] and [001] orientations. HREM image simulations based on the structural model obtained by direct methods are superimposed (Δf=-20nm, t=6nm).

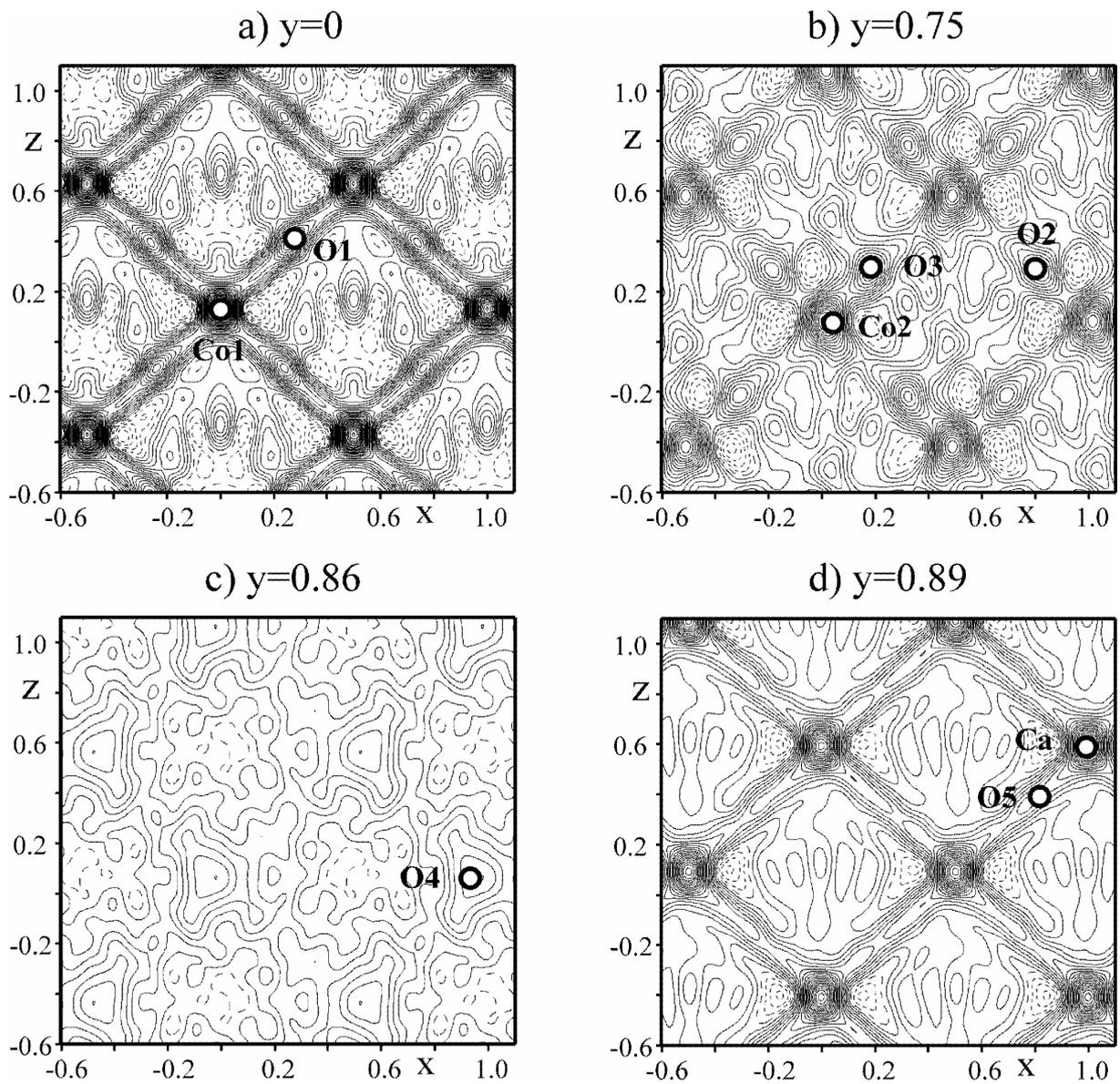

**Figure 8**: Sections for different y values of the electron density as obtained by inverse Fourier transform.

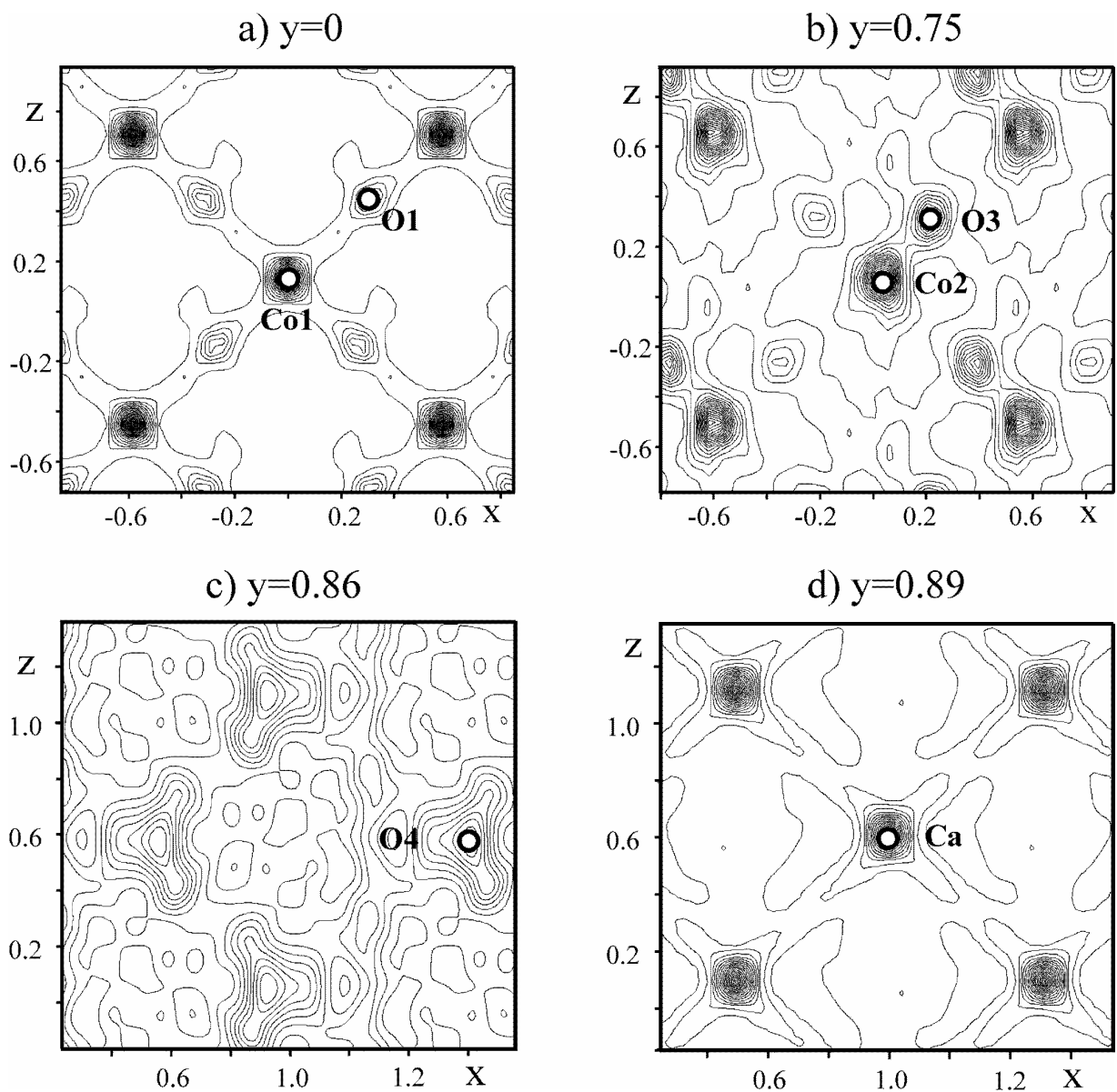

**Figure 9**: Sections for different y values (see Fig. 8) of the electron density calculated using the maximum entropy method.

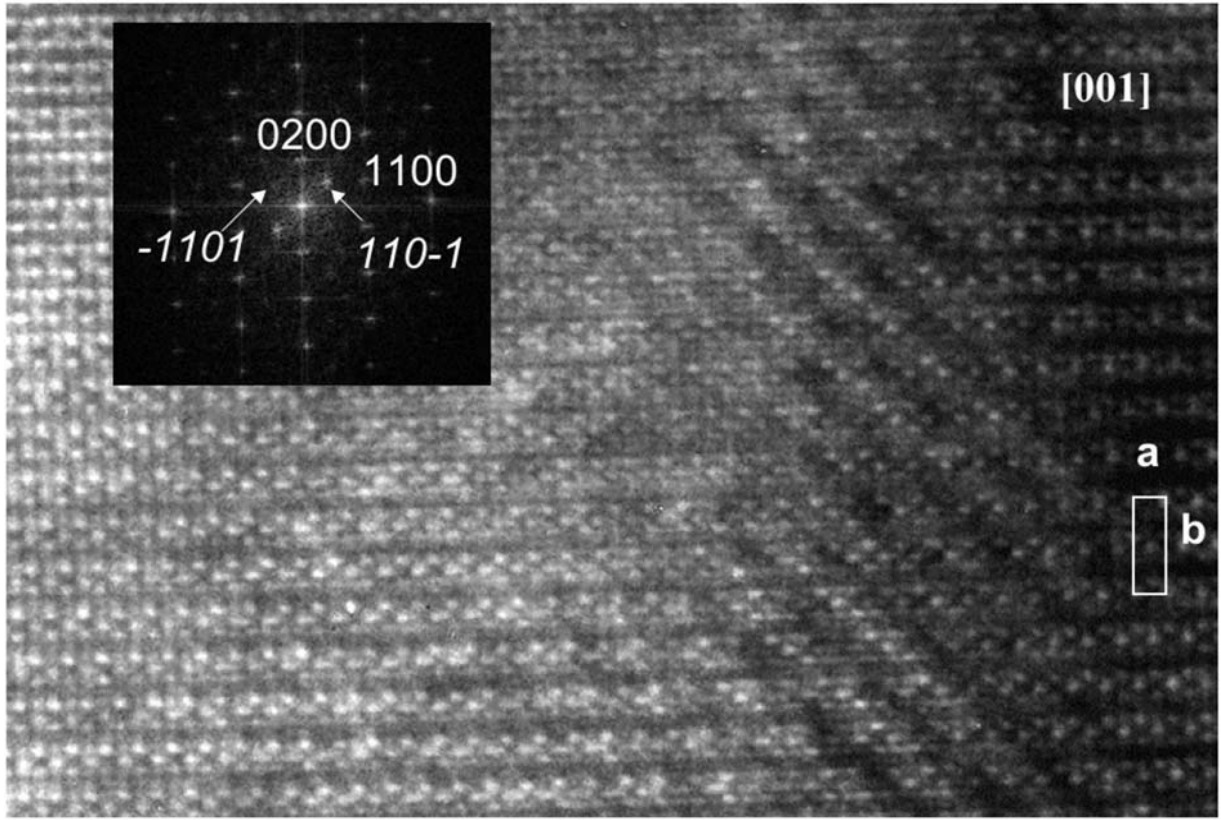

**Figure 10**: Example of nanodomain formed in [001]-oriented zones (the focus value is similar to that of Fig. 7c). The FT, in inset, evidences additional reflections with α≈0.56.